\documentclass[prd,showpacs,preprintnumbers,amsmath,amssymb,11pt]{revtex4}

\usepackage{graphics}
\usepackage[utf8]{inputenc}
\usepackage{epsfig}
\usepackage{subfigure}
\usepackage{dcolumn}
\usepackage{bm}
\usepackage{color}

\begin{document}
\title{More hidden heavy quarkonium molecules and their discovery  decay modes}

\author{Gang Li,$^{1,3}$\footnote{gli@mail.qfnu.edu.cn} Xiao Hai Liu,$^2$ and Zhu Zhou$^1$ }

\affiliation{ $^1$Department of Physics, Qufu Normal University, Qufu
273165, People's Republic of China}

\affiliation{$^2$Department of Physics and State Key Laboratory of Nuclear Physics and Technology, Peking University, Beijing 100871, People’s Republic of China}

\affiliation{ $^3$State Key Laboratory of Theoretical Physics, Institute of Theoretical Physics, Chinese Academy of Sciences, Beijing 100190, People’s Republic of China}


\begin{abstract}
To validate  the molecular description of the observed $Z_b(10610)/Z_b(10650)$ and $Z_c(3900)/Z_c(4025)$, it is valuable  to investigate their counterparts, denoted as $Z_{QV}^{(\prime)}$ in this work,  and the corresponding decay modes. In this work, we present an analysis of the $Z_{QV}^{(\prime)}$  using flavor symmetry. We also use the effective Lagrangian
based on the heavy quark symmetry to explore the rescattering mechanism and calculate the partial widths for the  isospin conserved channels $Z_{QV}^{(\prime)} \to \eta_Q V$. The predicted partial widths are of an order of MeV for $Z_{QV} \to \eta_Q V$, which correspond to branching ratios of the order of $10^{-2}\sim 10^{-1}$. For $Z_{QV}^\prime \to \eta_Q V$, the partial widths are a few hundreds of keV and the branching ratios are about $10^{-3}$. Future experimental measurements can test our predictions on the partial widths and thus examine the molecule description  of heavy quarkoniumlike exotic states.
\end{abstract}

\date{\today}
\maketitle

\section{Introduction}
\label{sec:introduction}
In the past few years, experiments have made great progress on the observations of $XYZ$ states and some of them cannot be  accommodated in the quark model as $Q{\bar Q}$ mesons~\cite{Beringer:1900zz}.
Among these states, charged charmoniumlike
and bottomoniumlike have attracted special attention due
to their four-quark nature~\cite{Brambilla:2010cs,Swanson:2006st,Eichten:2007qx,Voloshin:2007dx}.
In 2011, the Belle Collaboration reported two charged bottomoniumlike structures, $Z_b^{\pm}(10610)$ and $Z_b^{\pm}(10650)$, in the $\Upsilon(nS)\pi^{\pm}$ ($n=1,2,3$) and $h_b(mP)\pi^{\pm}$ ($m=1,2$) invariant mass spectra of $e^+ e^- \to \Upsilon(5S) \to \Upsilon(nS)\pi^+ \pi^-$ and $e^+ e^- \to \Upsilon(5S) \to h_b(mP) \pi^+\pi^-$~\cite{Collaboration:2011gja,Belle:2011aa}. The measured masses of $Z_b^{\pm}(10610)$ and $Z_b^{\pm}(10650)$ are slightly above the $B{\bar B}^*$ and $B^*{\bar B}^*$ thresholds, respectively. In 2013, the BESIII Collaboration observed a new charged state $Z_c^{\pm}(3900)$ in the $J/\psi \pi^{\pm}$ invariant mass spectrum of $Y(4260)\to J/\psi \pi^+\pi^-$~\cite{Ablikim:2013mio}. Later, this new charged charmoniumlike structure was also observed in the  $J/\psi \pi^{\pm}$ invariant mass spectrum by the Belle Collaboration~\cite{Liu:2013dau} and
confirmed by an analysis based on the CLEO data at the energy of $4.17$ GeV~\cite{Xiao:2013iha}. Another new charged structure, $Z_c^{\pm}(4025)$, was reported in the process $e^+e^- \to (D^*{\bar D}^*)^\pm \pi^{\mp}$ at ${\sqrt s} =4.26$ GeV by the BESIII Collaboration~\cite{Ablikim:2013emm}. Different from the other charmoniumlike and bottomoniumlike states, such as $X(3872)$, $Y(4260)$ ,etc., $Z_c^{(\prime)}$ and $Z_b^{(\prime)}$ are electric charged states and thus cannot be heavy quarkonium states. A charged combination could be formed by a state composed of four quarks, so these states may be ideal candidates for exotic hadrons beyond the conventional $Q {\bar Q}$ mesons.

The discoveries of these charged $Z_c^{(\prime)}$ and $Z_b^{(\prime)}$ states immediately initiated numerous studies of their structure, production, and decay mechanisms. Since the masses of the discovered states lie slightly above the meson-meson thresholds, it has been suggested that they are $S$-wave molecular states of heavy meson pair thresholds, i.e., $B^{(*)}{\bar B}^*$ and $D^{(*)}{\bar D}^*$~\cite{Sun:2011uh,Cleven:2011gp,Wang:2013cya,Bondar:2011ev,Cleven:2013sq}. Besides this explanation, these states also have been identified as tetraquark states based on the fact that these particles have a typical hadronic total width of a few tens of MeV~\cite{Ke:2012gm,Ali:2011ug,Braaten:2013boa,Faccini:2013lda}. Besides the spectrum study, the production and decay of $Z_c^{(\prime)}$ and $Z_b^{(\prime)}$ states have also been investigated extensively~\cite{Li:2012uc,Dong:2013iqa,Dong:2012hc,Guo:2013ufa,Guo:2014sca,Chen:2011zv,Chen:2013coa,Li:2012as,Chen:2011pv,Li:2014uia}.

On one hand, the molecular description can explain the  existing data on $Z_c^{(\prime)}/Z_b^{(\prime)}$; on the other hand, this interpretation has predicted  more counterpart states to be discovered. In this work,   we will investigate these new hidden heavy quarkonium states and more particularly the isospin conserved processes $Z_{QV}^{(\prime)} \to \eta_Q V$ which are of substantial importance in discovering  these new states.
As is well known, the intermediate meson loop transitions have been an important nonperturbative
transition mechanism in many processes, and their impact on the heavy quarkonium transitions has
been noticed for a long time~\cite{Lipkin:1986bi,Lipkin:1988tg,Moxhay:1988ri}. Recently, this mechanism has been applied to study $B$ decays~\cite{Cheng:2004ru,Lu:2005mx} and the production and decays of exotic states~\cite{Wang:2013cya,Bondar:2011ev,Cleven:2013sq,Chen:2011zv,Chen:2013coa,Li:2012as,Chen:2011pv,Li:2013yla,Guo:2010ak,Guo:2009wr,Liu:2013vfa,Guo:2013zbw,
Wang:2013hga,Voloshin:2013ez,Voloshin:2011qa,Guo:2010zk,Chen:2011pu,Chen:2012yr,Chen:2013bha}, and a global agreement with experimental data was obtained. Inspired by this agreement, we shall adopt the effective Lagrangian approach (ELA) to study the  decays  $Z_{QV}^{(\prime)} \to \eta_Q V$.

The rest of this paper is organized as follows. In Sec.~\ref{sec:molecules}, we present the possible molecular states composed of one pseudoscalar and one-vector ($P$-$V$) heavy mesons and two-vector ($V$-$V$) heavy mesons. In Sec.~\ref{sec:formula}, we will introduce the formulas for ELA. In Sec.~\ref{sec:results}, the numerical results are presented, and a brief summary is given in Sec.~\ref{sec:summary}.

\section{Hadronic Molecular States}
\label{sec:molecules}


\begin{table}[htb]
\begin{center}
\caption{Possible molecular states $Z_{QV}$ and $Z_{QV}^\prime$ with quantum numbers $J^{PC} = 1^{+-}$ for the neutral $Z_{QV}^{(\prime)}$.}\label{tab:molecule-1}
\begin{tabular}{c|c|c|c|c}
\hline \hline
$I,I_3$ & Label & States  & Label & States\\ \hline
$1,1$ & $Z_{c\rho}^+$ & $\frac {1} {\sqrt{2}} ({\bar D}^{*0} D^+ +{\bar D}^0 D^{*+})$ & $Z_{c\rho}^{\prime +} $ & ${\bar D}^{*0} D^{*+}$\\
$1,-1$ & $Z_{c\rho}^-$ & $ \frac {1} {\sqrt{2}} (D^{*0} D^- + {D}^0 D^{*-})$ & $Z_{c\rho}^{\prime -}$ & $D^{*0} D^{*-}$\\
$1,0$ &$Z_{c\rho}^0$ & $\frac {1} {2} [({\bar D}^{*0} D^0 -D^{*-} D^+ ) +({\bar D}^0  D^{*0} -D^- D^{*+})]$ &$Z_{c\rho}^{\prime 0}$ & $\frac {1}{\sqrt{2}} [({\bar D}^{*0} D^{*0} -D^{*-} D^{*+} )]$ \\
$0,0$ & $Z_{c\omega}^0$ & $\frac {1} {2} [({\bar D}^{*0} D^0 +D^{*-} D^+ ) +({\bar D}^0  D^{*0} + D^- D^{*+})]$ & $Z_{c\omega}^{\prime 0}$ & $\frac {1}{\sqrt{2}} [({\bar D}^{*0} D^{*0} +D^{*-} D^{*+} )]$\\
$0,0$ & $Z_{c\phi}^0$ & $\frac {1} {\sqrt{2}} ({D_s^{*+}} {D_s^-} + {D_s^{+}} {D_s^{*-}})$ & $Z_{c\phi}^{\prime 0}$ & ${D_s^{*+}} {D_s^{*-}}$\\
$\frac {1} {2}, \frac {1} {2}$ & $Z_{c{K^*}}^+$ &$\frac {1} {\sqrt{2}} ({\bar D}^{*0} D_s^+ +{\bar D}^0 D_s^{*+})$ & $Z_{c{K^*}}^+$ & ${\bar D}^{*0} D_s^{*+}$\\
$\frac {1} {2}, \frac {1} {2}$ & $Z_{c{K^*}}^-$ &$\frac {1} {\sqrt{2}} (D^{*0} D_s^- +D^0 D_s^{*-})$ & $Z_{c{K^*}}^-$ & $D^{*0} D_s^{*-}$\\
$\frac {1} {2}, -\frac {1} {2}$ & $Z_{cK^*}^0$ &$ \frac {1} {\sqrt{2}} ({D^{*-}} {D_s^+} + {D^{-}} {D_s^{*+}})$ & $Z_{cK^*}^{\prime 0}$ & ${D^{*-}} {D_s^{*+}}$\\
$\frac {1} {2}, \frac {1} {2}$ & $Z_{c{\bar K}^*}^0$ &$ \frac {1} {\sqrt{2}} ({D^{*+}} {D_s^-} + D^{+} D_s^{*-})$ & $Z_{c{\bar K}^*}^{\prime 0}$ & ${D^{*+}} D_s^{* -}$\\
\hline \hline
$I,I_3$ & Label & States  & Label & States\\ \hline
$1,1$ & $Z_{b\rho}^+$ & $\frac {1} {\sqrt{2}} (B^{*+} {\bar B}^0+B^+ {\bar B}^{*0})$ & $Z_{b\rho}^{\prime +} $ & $B^{*+} {\bar B}^{*0}$\\
$1,-1$ & $Z_{b\rho}^-$ & $\frac {1} {\sqrt{2}} (B^{*-} B^0 +B^-  B^{*0})$ & $Z_{b\rho}^{\prime -}$ & $B^{*-} B^{*0}$\\
$1,0$ &$Z_{b\rho}^0$ & $\frac {1} {2} [(B^{*+} B^- -B^{*0} {\bar B}^0) +(B^+  B^{*-} -B^0{\bar B}^{*0} )]$ &$Z_{b\rho}^{\prime 0}$ & $\frac {1} {\sqrt {2}} (B^{*+} B^{*-} -B^{*0} {\bar B}^{*0} )$ \\
$0,0$ & $Z_{b\omega}$ & $\frac {1} {2} [(B^{*+} B^- +B^{*0} {\bar B}^0) +(B^+  B^{*-} +B^0{\bar B}^{*0} )]$ & $Z_{b\omega}^\prime$ & $\frac {1} {\sqrt {2}} (B^{*+} B^{*-} +B^{*0} {\bar B}^{*0} )$\\
$0,0$ & $Z_{b\phi}$ & $\frac {1} {\sqrt{2}} ({B_s^{*}} {{\bar B}_s} + {B_s} {{\bar B}_s^{*}})$ & $Z_{b\phi}^\prime $ & ${B_s^{*}} {{\bar B}_s^{*}}$\\
$\frac {1} {2}, \frac {1} {2}$ & $Z_{b{K^*}}^+$ &$\frac {1} {\sqrt{2}} (B^{*+} {\bar B}_s + B^+ {\bar B}_s^{*})$ & $Z_{b{K^*}}^{\prime +}$ & $B^{*+} {\bar B}_s^{*}$\\
$\frac {1} {2}, -\frac {1} {2}$ & $Z_{b{K^*}}^-$ &$\frac {1} {\sqrt{2}} ({\bar B}^{*+} B_s + {\bar B}^+  B_s^{*})$ & $Z_{b{K^*}}^{\prime -}$ & ${\bar B}^{*+} B_s^{*}$\\
$\frac {1} {2}, -\frac {1} {2}$ & $Z_{bK^*}^0$ &$\frac {1} {\sqrt{2}} ({B^{*0}} {{\bar B}_s} + {B^{0}} {{\bar B}_s^{*}})$ & $Z_{bK^*}^{\prime 0}$ & ${B^{*0}} {\bar B}_s^{*}$\\
$\frac {1} {2}, \frac {1} {2}$ & $Z_{b{\bar K}^*}^0$ &$\frac {1} {\sqrt{2}} ({\bar B}^{*0} B_s + {\bar B}^0 B_s^{*})$ & $Z_{b{\bar K}^*}^{\prime 0}$ & ${\bar B}^{*0} B_s^{*}$\\
\hline \hline
\end{tabular}
\end{center}
\end{table}

For simplicity, we use $Z_{QV}^{(\prime)}$ ($Q=c,b$ and $V=\rho$, $\omega$, $\phi$ and $K^*$) to represent the states with the mesonic constituents $H^{(*)}{\bar H}^{(*)}$, where the heavy quark is denoted as $Q$ and $H^{(*)}{\bar H}^{(*)}$ represents
the relevant heavy (anti)meson.  The $(I, I_3)$ are the same as $V$.

In Table.~\ref{tab:molecule-1}, we collect the possible molecular states composed of one pseudoscalar and one vector ($P$-$V$) heavy mesons and two vector ($V$-$V$) heavy mesons. The quantum number is $J^{PC} = 1^{+-}$ for the neutral $Z_{QV}^{(\prime)}$.

\section{Decay Amplitudes}
\label{sec:formula}

\begin{figure}[ht]
\centering
\includegraphics[width=0.75\textwidth]{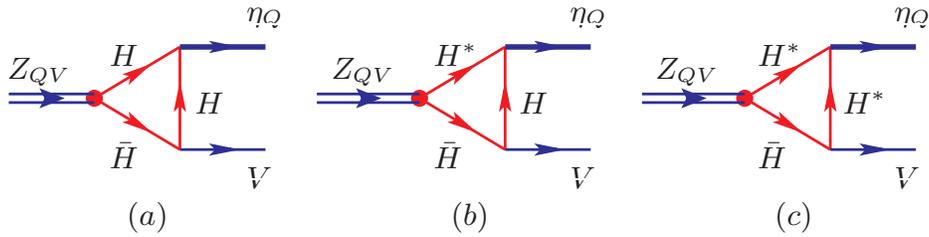}
\caption{The hadron-level diagrams for $Z_{QV} \to \eta_Q V$
via $H {\bar H}^{*} +c.c.$ intermediate heavy meson loops.}\label{fig:feyn-zQ-1}
\end{figure}

\begin{figure}[ht]
\centering
\includegraphics[width=0.5\textwidth]{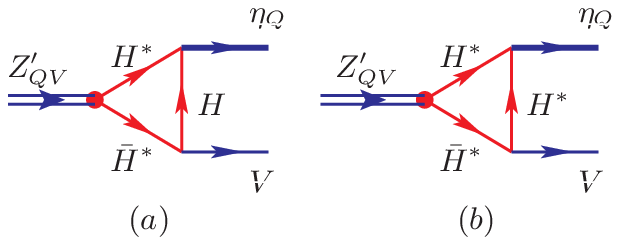}
\caption{The hadron-level diagrams for $Z_{QV}^{\prime} \to \eta_Q V$
via $H^* {\bar H}^*$ intermediate heavy meson loops.}\label{fig:feyn-zQ-2}
\end{figure}

Our calculation is based on the assumption that the $Z_{QV}^{(\prime)}$ are $S$-wave $H^{(*)}{\bar H}^{(*)}$ molecular states. The relevant Lagrangians for $Z_{QV}^{(\prime)}$ coupled to a pair of heavy mesons can be expressed as
\begin{eqnarray}
\mathcal{L}_{Z_{QV}, Z_{QV}^\prime}=
ig_{Z_{QV}^{\prime}} \varepsilon^{\mu \nu \alpha \beta}
{\bar H}^{*\dagger}_{\alpha} \partial_{\mu} Z_{QV \nu}^\prime H^{*\dagger}_{\beta} + g_{Z_QV}
( {\bar H}^{* \dagger }_{\mu}Z_{QV}^{\mu} H^\dagger + {\bar H}^\dagger Z_{QV}^{\mu} H^{* \dagger}_\mu) + H.c. \, , \label{eq:h4}
\end{eqnarray}
where $H$ and $H^*$ denote the pseudoscalar and vector heavy meson fields, respectively, i.e., $H^{(*)}=(D^{(*) 0},D^{ (*) +},D_s^{(*) +})$ and $( B^{(*) +}, B^{(*) 0}, B_s^{(*) 0})$.

With the experimental measurements for $BR(Z_{b\rho}^+ \to B^+{\bar B}^{*0}+{\bar B}^0 B^{*+})=(86.0\pm 3.6)\%$ and $BR(Z_{b\rho}^{\prime +} \to B^{*+}{\bar B}^{*0})=(73.4\pm 7.0)\%$~\cite{Adachi:2012cx}, we obtain
\begin{eqnarray} \label{eq:zb-coupling}
g_{Z_{b\rho}}= 13.10^{+0.83}_{-0.88} {\rm GeV}, \quad g_{Z_{b\rho}^\prime}=1.04^{+0.10}_{-0.10} \, .
\end{eqnarray}

If we assume that the total width of $Z_c$ and $Z_c^\prime$ are saturated by the decay $Z_{c\rho}^+ \to D^+ {\bar D}^{*0} + {\bar D}^0 D^{*+}$ and $Z_{c\rho}^{\prime +} \to D^{* +} {\bar D}^{*0}$, then the coupling constants are determined as follows,
\begin{eqnarray} \label{eq:zc-coupling}
g_{Z_{c\rho}}= 1.75^{+0.24}_{-0.25}  {\rm GeV} , \quad g_{Z_{c\rho}^\prime}=0.35^{+0.05}_{-0.06} \, .
\end{eqnarray}

The Lagrangian describing the interactions between $S$-wave heavy mesons and light vector mesons are as follows~\cite{Casalbuoni:1996pg,Cheng:2004ru}:
\begin{eqnarray}
{\cal L} &=&
ig_{H^*H^*\mathcal{V}} H^{*\nu\dagger}_i {\stackrel{\leftrightarrow}{\partial}}{\!_\mu} H^{*j}_\nu(\mathcal{V}^\mu)^i_j
+4if_{H^*H^*\mathcal{V}} H^{*\dagger}_{i\mu}(\partial^\mu \mathcal{V}^\nu-\partial^\nu
\mathcal{V}^\mu)^i_j H^{*j}_\nu  \nonumber \\
&&  -2f_{H^*H\mathcal{V}} \epsilon_{\mu\nu\alpha\beta}
(\partial^\mu \mathcal{V}^\nu)^i_j
(H_i^\dagger{\stackrel{\leftrightarrow}{\partial}}{\!^\alpha} H^{*\beta j}-H_i^{*\beta\dagger}{\stackrel{\leftrightarrow}{\partial}}{\!^\alpha} H^j) - ig_{HH\mathcal{V}} H_i^\dagger {\stackrel{\leftrightarrow}{\partial}}{\!_\mu} H^j(\mathcal{V}^\mu)^i_j
\, . \label{eq:LDDV}
\end{eqnarray}

In the chiral and heavy quark limits, the heavy meson couplings to the light vector meson have the following
relationships~\cite{Casalbuoni:1996pg,Cheng:2004ru},
\begin{eqnarray}
g_{HH{\cal V}}=g_{H^*H^*{\cal V}}=\frac{\beta
g_V}{\sqrt2}, \quad
f_{H^*H{\cal V}}&=&\frac{f_{H^*H^*{\cal V}}}{m_{H^*}}=\frac{\lambda g_V}{\sqrt2},
\end{eqnarray}
where $f_\pi = 132$ MeV is the pion decay constant, the parameters $g_V$ respect the relation $g_V = {m_\rho /
f_\pi}$~\cite{Casalbuoni:1996pg}. We take $\beta=0.9$, $\lambda = 0.56 \,
{\rm GeV}^{-1} $, and $g = 0.59$~\cite{Isola:2003fh}.

Based on the heavy quark symmetry~\cite{Colangelo:2003sa,Casalbuoni:1996pg}, the Lagrangians for the $S$-wave $\eta_Q$ couplings to $H$ and $H^*$ are
\begin{eqnarray}
\mathcal{L}= ig_{\eta_Q H^*H} H^{*\dagger\mu} (\partial_\mu \eta_Q H^\dagger -\eta_Q \partial_\mu H^\dagger) +g_{\eta_Q H^* H^*} \varepsilon^{\mu\nu\alpha\beta}
\partial_\mu H^{\dagger \ast}_\nu \partial_\alpha\eta_Q H^{*\dagger}_\beta  \, .
\label{eq:h1}
\end{eqnarray}
The following couplings are adopted in the numerical calculations:
\begin{eqnarray}
g_{\eta_Q H^* H}= 2g_Q\sqrt{m_{\eta_Q} m_{H^*}m_H}\, , \quad g_{\eta_Q H^* H^*} =2g_Q \frac {m_{H^*}} {\sqrt{m_{\eta_Q}}} \, ,
\end{eqnarray}
where $g_c=\sqrt{m_{J/\psi}}/(2m_{\cal D} f_{J/\psi})$, with $f_{J/\psi}= 405\pm 14$ MeV. $g_b=\sqrt{m_{\Upsilon(1S)}}/(2m_{\cal B} f_{\Upsilon(1S)})$ and $f_{\Upsilon(1S)}= 715.2$ MeV
is obtained via the experiment data~\cite{Beringer:1900zz}.
Based on the relevant Lagrangians given above, the loop transition amplitudes for the transitions in
Figs.~\ref{fig:feyn-zQ-1} and \ref{fig:feyn-zQ-2} can be expressed in a general form in the effective Lagrangian approach as follows:
 \begin{eqnarray}
 M_{fi}=\int \frac {d^4 q_2} {(2\pi)^4} \sum_{H^* \ \mbox{pol.}}
 \frac {V_1V_2V_3} {a_1 a_2 a_3}{\cal F}(m_{2}, q_2^2)
 \end{eqnarray}
where $V_i \ (i=1,2,3)$ are the vertex functions and $a_i = q_i^2-m_i^2
\ (i=1,2,3)$ are the denominators of the intermediate heavy meson
propagators. For example, in Fig.~\ref{fig:feyn-zQ-1} (a), $V_i$ ($i=1, 2, 3$) are the vertex functions for the $Z_{QV}$, $\eta_Q$, and $V$ mesons, respectively. $V_i$ ($i=1, 2, 3$) are the denominators for the intermediate $H^{*}$, ${\bar H}^{*}$, $H$, respectively. We adopt a monopole form factor,
\begin{eqnarray}\label{ELA-form-factor}
{\cal F}(m_{2}, q_2^2) \equiv \frac
{\Lambda^2-m_{2}^2} {\Lambda^2-q_2^2},
\end{eqnarray}
where $\Lambda_i\equiv m_i+\alpha\Lambda_{\rm QCD}$ and the QCD energy scale $\Lambda_{\rm QCD} = 220$ MeV. This parameter scheme has been applied extensively in other works~\cite{Li:2012as,Guo:2010ak,Cheng:2004ru,Liu:2009vv,Liu:2010um}. This form factor is supposed to offset the off-shell effects of the exchanged mesons~\cite{Cheng:2004ru,Li:1996yn,Locher:1993cc,Li:1996cj} and the form factor parameter should be determined by experimental information. The explicit expression of the transition amplitudes can be found in the Appendix~\ref{appendix-A}.

\section{Numerical results}
\label{sec:results}

\begin{table}
\begin{center}
\caption{A summary of meson masses (in units
of MeV) adopted in the calculation. }\label{tab:meson-mass}
\begin{tabular}{ccccccccc}
\hline\hline
States       & $Z_{b\rho}^+$ & $Z_{b\rho}^{\prime +}$  & $\eta_b$ & $B^*$  & $B$ & $B_s^*$ & $B_s$ \\
Mass  & $10607.2$ & $10652.2$ & $9391.0$ & $5325.2$ & $5279.25$ & $5415.4$ & $5366.77$ \\
\hline\hline
States       & $Z_{c\rho}^+$ & $Z_{c\rho}^{\prime +}$ & $\eta_c$ & $D^*$  & $D$ & $D_s^*$ & $D_s$ \\
Mass  & $3891.5$ & $4023.0$ &$2981.0$ &$2010.28$& $1869.62$ & $2112.3$ & $1968.49$ \\
\hline\hline
\end{tabular}
\end{center}
\end{table}

\begin{table}[htb]
\begin{center}
\caption{Predicted partial widths (in unit of MeV) of $Z_{cV} \to \eta_c V$.  The  parameter in the form factor is chosen as $\alpha =2.0$ and $\alpha =3.0$. }\label{tab:results-1}
\begin{tabular}{ccccccccccc}
\hline
& \multicolumn{4}{c} {$\alpha=2.0$} & \multicolumn{4}{c} {$\alpha=3.0$} \\ \hline
Binding energy & $-20$ MeV & $-10$ MeV & $-5$ MeV &$-1$ MeV  & $-20$ MeV & $-10$ MeV & $-5$ MeV &$-1$ MeV \\ \hline
$Z_{c\rho}^+ \to \eta_c \rho^+$  & $0.44$ & $0.42$ & $0.41$ & $0.41$  & $0.76$ & $0.73$ & $0.71$ & $0.70$     \\ \hline
$Z_{c\rho}^0 \to \eta_c \rho^0$  & $0.44$ & $0.42$ & $0.41$ & $0.37$  & $0.76$ & $0.73$ & $0.72$ & $0.64$  \\ \hline
$Z_{c\omega}^0 \to \eta_c \omega$ & $0.43$& $0.41$ & $0.40$ & $0.35$  & $0.75$  & $0.72$& $0.70$ & $0.63$     \\ \hline
$Z_{c\phi}^0 \to \eta_c \phi$  & $0.09$ & $0.08$  & $0.08$ & $0.08$      & $0.15$ & $0.15$ & $0.14$ & $0.14$  \\ \hline
$Z_{cK^{*}}^+ \to \eta_c K^{*+}$ & $0.41$ & $0.39$ & $0.38$ & $0.37$     & $0.71$ & $0.68$ & $0.67$ & $0.66$   \\ \hline
$Z_{c K^{*}}^0 \to \eta_c K^{*0}$ & $0.40$ & $0.39$ & $0.38$ & $0.35$   & $0.71$ & $0.68$ & $0.67$ & $0.63$  \\ \hline
\end{tabular}
\end{center}
\end{table}

\begin{table}[htb]
\begin{center}
\caption{Predicted partial widths (in unit of keV) of $Z_{cV}^\prime \to \eta_c V$.  The  parameter in the form factor is chosen as $\alpha =2.0$ and $\alpha =3.0$. }\label{tab:results-2}
\begin{tabular}{ccccccccc}
\hline
& \multicolumn{4}{c} {$\alpha=2.0$} & \multicolumn{4}{c} {$\alpha=3.0$} \\ \hline
Binding energy & $-20$ MeV & $-10$ MeV & $-5$ MeV &$-1$ MeV  & $-20$ MeV & $-10$ MeV & $-5$ MeV &$-1$ MeV \\ \hline
$Z_{c\rho}^{\prime +} \to \eta_c \rho^+$  & $12.43$ & $11.26$ & $10.69$ &$10.25$   & $22.88$ & $20.74$ & $19.72$ & $18.91$  \\ \hline
$Z_{c\rho}^{\prime 0} \to \eta_c \rho^0$   & $12.51$ & $11.32$ & $10.73$ & $9.37$  & $23.01$ & $20.84$ & $19.77$ & $17.45$   \\ \hline
$Z_{c\omega}^{\prime 0} \to \eta_c \omega$ & $11.81$ & $10.67$ & $10.09$ &$8.81$   & $21.76$ & $19.66$ & $18.62$ & $16.44$    \\ \hline
$Z_{c\phi}^{\prime 0} \to \eta_c \phi$ & $1.95$ & $1.74$ & $1.63$ & $1.55$   & $3.65$ & $3.26$ & $3.07$ & $2.93$     \\ \hline
$Z_{cK^*}^{\prime +}\to \eta_c K^{*+}$ & $10.44$ & $9.39$ & $8.89$ &$8.49$      & $19.29$ & $17.38$ & $16.46$ & $15.74$    \\ \hline
$Z_{cK^*}^{\prime 0} \to \eta_c K^{*0}$ & $10.30$ & $9.26$ & $8.76$ &$8.37$   & $19.05$ & $17.16$ & $16.24$ & $15.53$   \\ \hline
\end{tabular}
\end{center}
\end{table}

\begin{table}[htb]
\begin{center}
\caption{Predicted partial widths (in units of MeV) of $Z_{bV} \to \eta_b V$.  The  parameter in the form factor is chosen as $\alpha =2.0$ and $\alpha =3.0$. }\label{tab:results-3}
\begin{tabular}{ccccccccc}
\hline
& \multicolumn{4}{c} {$\alpha=2.0$} & \multicolumn{4}{c} {$\alpha=3.0$} \\ \hline
Binding energy & $-20$ MeV & $-10$ MeV & $-5$ MeV &$-1$ MeV  & $-20$ MeV & $-10$ MeV & $-5$ MeV &$-1$ MeV \\ \hline
$Z_{b\rho}^+ \to \eta_b \rho^+$   & $1.97$ & $1.93$ & $1.91$ & $1.89$ & $3.22$ & $3.16$ & $3.13$ & $3.11$ \\ \hline
$Z_{b\rho}^0 \to \eta_b \rho^0$   & $1.97$ & $1.93$ & $1.91$ & $1.89$ & $3.22$ & $3.16$ & $3.13$ & $3.11$  \\ \hline
$Z_{b\omega}^0 \to \eta_b \omega$ & $1.94$ & $1.91$ & $1.89$ & $1.87$ & $3.19$ & $3.13$ & $3.11$ & $3.08$ \\ \hline
$Z_{b\phi}^0 \to \eta_b \phi$     & $0.43$ & $0.42$ & $0.42$ & $0.41$ & $0.74$ & $0.73$ & $0.72$ & $0.72$ \\ \hline
$Z_{bK^*}^+ \to \eta_b K^{*+}$    & $1.88$ & $1.84$ & $1.82$ & $1.76$ & $3.15$ & $3.10$ & $3.07$ & $2.97$ \\ \hline
$Z_{bK^*}^0 \to \eta_b K^{*0}$    & $1.87$ & $1.83$ & $1.81$ & $1.75$ & $3.14$ & $3.08$ & $3.05$ & $2.95$ \\ \hline
\end{tabular}
\end{center}
\end{table}

\begin{table}[htb]
\begin{center}
\caption{Predicted partial widths (in units of keV) of $Z_{bV}^\prime \to \eta_b V$.  The  parameter in the form factor is chosen as $\alpha =2.0$ and $\alpha =3.0$. }\label{tab:results-4}
\begin{tabular}{ccccccccc}
\hline
& \multicolumn{4}{c} {$\alpha=2.0$} & \multicolumn{4}{c} {$\alpha=3.0$} \\ \hline
Binding energy & $-20$ MeV & $-10$ MeV & $-5$ MeV &$-1$ MeV  & $-20$ MeV & $-10$ MeV & $-5$ MeV &$-1$ MeV \\ \hline
$Z_{b\rho}^{\prime +} \to \eta_c \rho^+$   & $105.01$ & $99.46$ & $96.73$ & $94.58$  & $196.68$ & $186.11$ & $180.93$  & $176.84$ \\ \hline
$Z_{b\rho}^{\prime 0} \to \eta_c \rho^0$   & $105.01$ & $99.46$ & $96.73$ & $94.58$  & $196.68$ & $186.11$ & $180.93$  & $176.84$  \\ \hline
$Z_{b\omega}^{\prime 0} \to \eta_b \omega$ & $101.52$ & $96.06$ & $93.39$ & $91.27$  & $190.10$ & $179.71$ & $174.63$  & $170.61$  \\ \hline
$Z_{b\phi}^{\prime 0} \to \eta_b \phi$     & $17.98$  & $16.84$ & $16.29$ & $15.85$  & $33.95$  & $31.78$  & $30.72$   & $29.89$ \\ \hline
$Z_{b K^*}^{\prime +} \to \eta_b K^{*+}$   & $90.79$  & $85.64$ & $83.11$ & $81.12$  & $170.51$ & $160.68$ & $155.87$  & $152.07$  \\ \hline
$Z_{bK^*}^{\prime 0} \to \eta_b K^{*0}$    & $88.88$  & $83.78$ & $81.28$ & $79.31$  & $166.91$ & $157.19$ & $152.43$  & $148.67$   \\ \hline
\end{tabular}
\end{center}
\end{table}

The molecule is the pole of the S matrix. As a result, it could be bound state (on a physical sheet below the threshold of constituent particles), virtual state (on an unphysical sheet below the threshold), or resonance (on an unphysical sheet above the threshold)~\cite{Weinberg:1962hj,Weinberg:1963zza,Weinberg:1965zz,Badalian:1981xj,Guo:2008zg,Cleven:2011gp}. In our case, we assume these exotic states studied here are resonances and can decay into their constituent particles. In Table~\ref{tab:meson-mass}, we list the meson masses involved in our calculation. Based on masses of the discovered states, we can get that the binding energy defined as $E_Z= m_{H^{(*)}}+m_{{\bar H}^{(*)}}-m_Z$ are about $(-12\sim -2)$ MeV. These measurements have provided a range for the binding
energy and we will choose a few illustrative values, $E_Z=(-20,-10,-5,-1  )$ MeV, in the following. Choosing two values for the cutoff parameter $\alpha$, we have predicted the partial decay widths, and the numerical results are collected in Tables~\ref{tab:results-1}-\ref{tab:results-4}.

As can be seen in Table~\ref{tab:results-1}, the predicted partial widths of $Z_{cV} \to \eta_c V$ are less sensitive to the cutoff parameter $\alpha$ and the binding energy $E_Z$. The partial widths are about $1$ MeV, except for $Z_{c\phi} \to \eta_c\phi$. It is noteworthy to recall that the experimental measurements for $\Gamma(Z_{c\rho})$ is $39.2 \pm 10.5$ MeV~\cite{Ablikim:2013mio,Liu:2013dau,Xiao:2013iha}. If these $Z_{cV}$ have similar widths, our results would indicate a sizable branching fractions, at least $10^{-2}$, for these decays. The partial width is only about $0.1$ MeV for $Z_{c\phi} \to \eta_c\phi$, which is because the phase space is much smaller with the $E_Z$ values considered in the calculations.

In Table~\ref{tab:results-2}, we list the partial widths of $Z_{c V}^\prime \to \eta_c V$. The behavior is similar to that of $Z_{cV} \to \eta_c V$. The predicted partial widths are about several keV  for $Z_{c\phi}^\prime \to \eta_c\phi$ and tens of keV for other $Z_{cV}^\prime \to \eta_c V$. If we assume that the $Z_{cV}^\prime$ have a width similar to $Z_c(4020)$, i.e., $9.7\pm 3.2$ MeV, the corresponding branching ratios are about $10^{-3}$, which is about $1$-$2$ orders of magnitude smaller than that of  $Z_{c V} \to \eta_c V$. As shown in Figs.~\ref{fig:feyn-zQ-1} and \ref{fig:feyn-zQ-2}, there are three kinds of diagrams for $Z_{cV}\to \eta_c V$, while there are only two kinds of diagrams for $Z_{cV}^\prime \to \eta_c V$. Of course, there are still uncertainties coming from the coupling constants and off-shell effects arising from the exchanged particles of the loops, and the cutoff parameter can also be different in different decays.
The numerical results would be lacking in accuracy and we expect to see experimental measurements on open charmed pair decays and the $\eta_c V$ decays in the near future.

The calculated partial widths of $Z_{bV} \to \eta_b V$ and $Z_{bV}^\prime \to \eta_b V$ are presented in Tables~\ref{tab:results-3} and \ref{tab:results-4}, respectively. The behaviors are similar to $Z_{cV}^{(\prime)} \to \eta_c V$. The predicted widths of $Z_{bQ} \to \eta_b V$ are about a few MeV. The experimental measured total width of $Z_{b\rho}$ is $(18.4\pm 2.4)$ MeV, taking into account the fact that these $Z_{bV}$ should have similar total widths and that the corresponding branching ratios can reach up to $20\%$.
The results for $Z_{b\phi}^0 \to \eta_b \phi$ are less than $1$ MeV, which is due to the suppressed phase space.
For most $Z_{bV}^\prime \to \eta_b V$, the partial widths are about $100$-$200$ keV, which corresponds to branching ratios of $10^{-2}$. Future experimental measurements can test our predictions and thus examine the properties of heavy quarkonium molecules.
\section{Summary}
\label{sec:summary}

In this work, we have investigated the partners of the observed $Z_b(10610)/Z_b(10650)$ and $Z_c(3900)/Z_c(4025)$ states under the molecular framework. We have explored the isospin conserved decays $Z_{QV}/Z_{QV}^{\prime } \to \eta_Q V$, respectively, via intermediate heavy meson loops. In this calculation, $Z_{QV}$ and $Z_{QV}^{\prime }$ are assumed to be $H {\bar H}^* + c.c.$  and $H^*{\bar H}^*$ molecular states, respectively. The quantum numbers of the neutral partners of these two resonances are fixed to be $J^{PC} = 1^{+-}$.

For these decays, our results show that the $\alpha$ dependence of the partial widths are not less sensitive to some extent. The binding energy dependence is also quite stable. The predicted partial widths are of order of MeV for $Z_{QV} \to \eta_Q V$, which corresponds to the branching ratios of an order $10^{-2}\sim 10^{-1}$. For $Z_{QV}^\prime \to \eta_Q V$, the partial widths are a few hundreds of keV and the branching ratios are about $10^{-3}$.
The results show some evidence of the universality of the molecular description of the observed $Z_c^{(\prime)}$ and $Z_b^{(\prime)}$, and they indicate that the intermediate heavy meson loops may be an important transition mechanism in the decays of the discovered exotic states, especially the initial states that are close to the two particle thresholds. We expect experiments to search for these $Z_{QV}^{(\prime)}$, the decays of $Z_{QV}^{(\prime)} \to \eta_Q V$, which will help us to test the universality of the molecular description of the $Z_{QV}^{(\prime)}$ and their transition mechanism deeply.

\section*{Acknowledgements}

The authors thanks Q. Wang and W. Wang for useful
discussions. This work is supported, in part, by the National
Natural Science Foundation of China (Grant No. 11275113) and the China Postdoctoral Science Foundation (Grant No. 2013M530461).

\begin{appendix}

\label{appendix-A}
\section{The Transition Amplitude in ELA}

In the following, we give the transition amplitudes for the
intermediate heavy meson loops listed in Figs.~\ref{fig:feyn-zQ-1} and
\ref{fig:feyn-zQ-2} in the framework of the ELA. $p_1$, $p_2$, and $p_3$ are the four-vector momenta for initial states, final heavy quarkonium $\eta_Q$, and final light vector mesons $V$, respectively. $q_1$, $q_2$ and $q_3$ are the four-vector momenta for the intermediate heavy mesons. $\varepsilon_1$ and $\varepsilon_3$ are the polarization vectors for initial state and final light vector mesons, respectively.

(i) $Z_{QV}^+ \to \eta_Q V$
\begin{eqnarray}
M_{HH^* [H^*]}&=& \int \frac {d^4q_2} {(2\pi)^4}[g_{Z_{QV}}
\varepsilon_{1\mu} ] [g_{\eta_Q H^*H}(p_2+q_1)_\rho ] [2g_{H^*H^*V} q_{2\theta} \varepsilon_3^{* \theta} g_{\phi\kappa} + 4f_{H^*H^*V} (p_{3}^{\theta}\varepsilon_{3\phi}^* - p_{3\phi}\varepsilon_{3}^{*\theta})g_{\kappa \theta}]
\nonumber \\
&& \times \frac {i} {q_1^2-m_1^2}  \frac {i(-g^{\rho\phi}
+q_2^\rho q_2^\phi/m_2^2)} {q_2^2-m_2^2}  \frac {i(-g^{\mu\kappa}
+q_3^\mu q_3^\kappa/m_2^2)} {q_3^2-m_3^2} {\cal F}(m_{2}, q_2^2), \nonumber \\
M_{H^*H [H^*]} &=& \int \frac {d^4q_2} {(2\pi)^4}[g_{Z_{QV}}
\varepsilon_{1\mu}] [g_{\eta_Q H^*H} (p_2-q_2)_\rho] [g_{HH V} (q_2-q_3)_\lambda \varepsilon_3^{*\lambda}] \nonumber \\
&& \times \frac {i(-g^{\mu\rho} +q_1^\mu q_1^\rho/m_1^2)}
{q_1^2-m_1^2}  \frac {i} {q_2^2-m_2^2}  \frac {i} {q_3^2-m_3^2} {\cal F}(m_{2}, q_2^2),  \nonumber \\
M_{H^*H^* [H]} &=& \int \frac {d^4q_2} {(2\pi)^4}[g_{Z_{QV}}
\varepsilon_{1\mu}] [g_{\eta_Q H^*H^*} \varepsilon_{\rho \sigma \xi\tau}q_2^\rho p_2^\tau ] [g_{H^*HV}\varepsilon_{\lambda \theta \phi\kappa} p_3^\lambda \varepsilon_3^{* \theta} (q_2-q_3)^\phi] \nonumber \\
&& \times \frac {i(-g^{\mu\xi} +q_1^\mu q_1^\xi/m_1^2)}
{q_1^2-m_1^2}  \frac {i(-g^{\sigma\kappa}
+q_2^\sigma q_2^\kappa/m_2^2)} {q_2^2-m_2^2}  \frac {i} {q_3^2-m_3^2} {\cal F}(m_{2}, q_2^2) \, .
\end{eqnarray}

(ii) $Z_{QV}^{\prime + } \to \eta_Q V$
\begin{eqnarray}
M_{H^*H^* [H]} &=& \int \frac {d^4q_2} {(2\pi)^4}[g_{Z_{QV}^{(\prime)} H^*H^*} \varepsilon_{\mu\nu\alpha\beta} p_1^\mu\varepsilon_{1}^\nu ] [g_{\eta_Q H^*H} (p_2-q_2)_\lambda] [g_{H^*HV}\varepsilon_{\theta\xi\tau\kappa} p_3^\theta\varepsilon_3^{* \xi} (q_3-q_2)^\tau] \nonumber \\
&& \times \frac {i(-g^{\alpha\lambda} +q_1^\alpha q_1^\lambda/m_1^2)} {q_1^2-m_1^2}  \frac {i} {q_2^2-m_2^2}  \frac {i(-g^{\beta\kappa} +q_3^\beta q_3^\kappa/m_3^2)} {q_3^2-m_3^2} {\cal F}(m_{2}, q_2^2),  \nonumber \\
M_{H^*H^* [H^*]} &=& \int \frac {d^4q_2} {(2\pi)^4}[g_{Z_{QV}^{(\prime)} H^*H^*} \varepsilon_{\mu\nu\alpha\beta} p_1^\mu\varepsilon_{1}^\nu ] [g_{\eta_Q H^*H^*} \varepsilon_{\lambda\sigma\xi\tau}q_2^\lambda p_2^\tau ] \nonumber \\ && \times [2g_{H^*H^*V} q_{2\theta} \varepsilon_3^{*\theta} g_{\phi\kappa} + 4f_{H^*H^*V} (p_{3}^{\theta}\varepsilon_{3\phi}^* - p_{3\phi}\varepsilon_{3}^{*\theta} )g_{\kappa \theta}] \nonumber \\
&& \times \frac {i(-g^{\alpha\xi} +q_1^\alpha q_1^\xi/m_1^2)}
{q_1^2-m_1^2}  \frac {i(-g^{\phi\sigma} +q_2^\phi q_2^\sigma/m_2^2)}
{q_2^2-m_2^2}  \frac {i(-g^{\beta\kappa} +q_3^\beta q_3^\kappa/m_3^2)}
{q_3^2-m_3^2} {\cal F}(m_{2}, q_2^2) \, .
\end{eqnarray}
\end{appendix}


\begin{thebibliography}{0}
\bibitem{Beringer:1900zz}
  J.~Beringer {\it et al.}  [Particle Data Group Collaboration],
  Phys.\ Rev.\ D {\bf 86}, 010001 (2012).

\bibitem{Brambilla:2010cs}
  N.~Brambilla, S.~Eidelman, B.~K.~Heltsley, R.~Vogt, G.~T.~Bodwin, E.~Eichten, A.~D.~Frawley and A.~B.~Meyer {\it et al.},
  Eur.\ Phys.\ J.\ C {\bf 71}, 1534 (2011)  [arXiv:1010.5827 [hep-ph]].

\bibitem{Swanson:2006st}
  E.~S.~Swanson,
  Phys.\ Rept.\  {\bf 429}, 243 (2006)  [hep-ph/0601110].

\bibitem{Eichten:2007qx}
  E.~Eichten, S.~Godfrey, H.~Mahlke and J.~L.~Rosner,
  Rev.\ Mod.\ Phys.\  {\bf 80}, 1161 (2008)  [hep-ph/0701208].

\bibitem{Voloshin:2007dx}
  M.~B.~Voloshin,
  Prog.\ Part.\ Nucl.\ Phys.\  {\bf 61}, 455 (2008)  [arXiv:0711.4556 [hep-ph]].

\bibitem{Collaboration:2011gja}
  I.~Adachi [Belle Collaboration],
  arXiv:1105.4583 [hep-ex].

\bibitem{Belle:2011aa}
  A.~Bondar {\it et al.}  [Belle Collaboration],
  Phys.\ Rev.\ Lett.\  {\bf 108}, 122001 (2012)  [arXiv:1110.2251 [hep-ex]].


\bibitem{Ablikim:2013mio}
  M.~Ablikim {\it et al.}  [BESIII Collaboration],
   Phys.\ Rev.\ Lett.\  {\bf 110}, 252001 (2013)  [arXiv:1303.5949 [hep-ex]].

\bibitem{Liu:2013dau}
  Z.~Q.~Liu {\it et al.}  [Belle Collaboration],
  Phys.\ Rev.\ Lett.\  {\bf 110}, 252002 (2013)  [arXiv:1304.0121 [hep-ex]].

\bibitem{Xiao:2013iha}
  T.~Xiao, S.~Dobbs, A.~Tomaradze and K.~K.~Seth,
  Phys.\ Lett.\ B {\bf 727}, 366 (2013)
  [arXiv:1304.3036 [hep-ex]].

\bibitem{Ablikim:2013emm}
  M.~Ablikim {\it et al.}  [BESIII Collaboration],
  Phys.\ Rev.\ Lett.\  {\bf 112}, 132001 (2014)
  [arXiv:1308.2760 [hep-ex]].

\bibitem{Sun:2011uh}
  Z.~-F.~Sun, J.~He, X.~Liu, Z.~-G.~Luo and S.~-L.~Zhu,
  Phys.\ Rev.\ D {\bf 84}, 054002 (2011)  [arXiv:1106.2968 [hep-ph]].

\bibitem{Cleven:2011gp}
  M.~Cleven, F.~-K.~Guo, C.~Hanhart and U.~-G.~Mei{\ss}ner,
  Eur.\ Phys.\ J.\ A {\bf 47}, 120 (2011)  [arXiv:1107.0254 [hep-ph]].

\bibitem{Wang:2013cya}
  Q.~Wang, C.~Hanhart and Q.~Zhao,
  Phys.\ Rev.\ Lett.\  {\bf 111}, 132003 (2013)  [arXiv:1303.6355 [hep-ph]].

\bibitem{Bondar:2011ev}
  A.~E.~Bondar, A.~Garmash, A.~I.~Milstein, R.~Mizuk and M.~B.~Voloshin,
  Phys.\ Rev.\ D {\bf 84}, 054010 (2011)  [arXiv:1105.4473 [hep-ph]].

\bibitem{Cleven:2013sq}
  M.~Cleven, Q.~Wang, F.~-K.~Guo, C.~Hanhart, U.~-G.~Mei{\ss}ner and Q.~Zhao,
  Phys.\ Rev.\ D {\bf 87}, no. 7, 074006 (2013)
  [arXiv:1301.6461 [hep-ph]].

\bibitem{Ke:2012gm}
  H.~-W.~Ke, X.~-Q.~Li, Y.~-L.~Shi, G.~-L.~Wang and X.~-H.~Yuan,
  JHEP {\bf 1204}, 056 (2012)  [arXiv:1202.2178 [hep-ph]].

\bibitem{Ali:2011ug}
  A.~Ali, C.~Hambrock and W.~Wang,
  Phys.\ Rev.\ D {\bf 85}, 054011 (2012)  [arXiv:1110.1333 [hep-ph]].

\bibitem{Braaten:2013boa}
  E.~Braaten,
  Phys.\ Rev.\ Lett.\  {\bf 111}, 162003 (2013)
  [arXiv:1305.6905 [hep-ph]].

\bibitem{Faccini:2013lda}
  L.~Maiani, V.~Riquer, R.~Faccini, F.~Piccinini, A.~Pilloni and A.~D.~Polosa,
  Phys.\ Rev.\ D {\bf 87}, no. 11, 111102 (2013)  [arXiv:1303.6857 [hep-ph]].

\bibitem{Li:2012uc}
  X.~Li and M.~B.~Voloshin,
  Phys.\ Rev.\ D {\bf 86}, 077502 (2012)  [arXiv:1207.2425 [hep-ph]].

\bibitem{Dong:2013iqa}
  Y.~Dong, A.~Faessler, T.~Gutsche and V.~E.~Lyubovitskij,
  Phys.\ Rev.\ D {\bf 88}, 014030 (2013)  [arXiv:1306.0824 [hep-ph]].

\bibitem{Dong:2012hc}
  Y.~Dong, A.~Faessler, T.~Gutsche and V.~E.~Lyubovitskij,
  J.\ Phys.\ G {\bf 40}, 015002 (2013)  [arXiv:1203.1894 [hep-ph]].

\bibitem{Guo:2013ufa}
  F.~-K.~Guo, U.~-G.~Mei{\ss}ner and W.~Wang,
  Commun.\ Theor.\ Phys.\  {\bf 61}, 354 (2014)
  [arXiv:1308.0193 [hep-ph]].

\bibitem{Guo:2014sca}
  F.~-K.~Guo, U.~-G.~Mei{\ss}ner and W.~Wang,
  arXiv:1402.6236 [hep-ph].

\bibitem{Li:2014uia}
  G.~Li and W.~Wang,
  Phys.\ Lett.\ B {\bf 733}, 100 (2014)
  [arXiv:1402.6463 [hep-ph]].

\bibitem{Chen:2011zv}
  D.~-Y.~Chen, X.~Liu and S.~-L.~Zhu,
  Phys.\ Rev.\ D {\bf 84}, 074016 (2011)  [arXiv:1105.5193 [hep-ph]].

\bibitem{Chen:2013coa}
  D.~-Y.~Chen, X.~Liu and T.~Matsuki,
  Phys.\ Rev.\ D {\bf 88}, 036008 (2013)  [arXiv:1304.5845 [hep-ph]].

\bibitem{Li:2012as}
  G.~Li, F.~-l.~Shao, C.~-W.~Zhao and Q.~Zhao,
  Phys.\ Rev.\ D {\bf 87}, no. 3, 034020 (2013)
  [arXiv:1212.3784 [hep-ph]].

\bibitem{Chen:2011pv}
  D.~-Y.~Chen and X.~Liu,
  Phys.\ Rev.\ D {\bf 84}, 094003 (2011)
  [arXiv:1106.3798 [hep-ph]].

\bibitem{Lipkin:1986bi}
  H.~J.~Lipkin,
  Nucl.\ Phys.\ B {\bf 291}, 720 (1987).


\bibitem{Lipkin:1988tg}
  H.~J.~Lipkin and S.~F.~Tuan,
  Phys.\ Lett.\ B {\bf 206}, 349 (1988).


\bibitem{Moxhay:1988ri}
  P.~Moxhay,
  Phys.\ Rev.\ D {\bf 39}, 3497 (1989).


\bibitem{Cheng:2004ru}
  H.~Y.~Cheng, C.~K.~Chua and A.~Soni,
  Phys.\ Rev.\  D {\bf 71}, 014030 (2005)
  [arXiv:hep-ph/0409317].

\bibitem{Lu:2005mx}
  C.~-D.~Lu, Y.~-L.~Shen and W.~Wang,
  Phys.\ Rev.\ D {\bf 73}, 034005 (2006)
  [hep-ph/0511255].

\bibitem{Li:2013yla}
  G.~Li and X.~-H.~Liu,
  Phys.\ Rev.\ D {\bf 88},  094008 (2013)
  [arXiv:1307.2622 [hep-ph]].

\bibitem{Guo:2010ak}
  F.~K.~Guo, C.~Hanhart, G.~Li, U.~G.~Meissner and Q.~Zhao,
  Phys.\ Rev.\  D {\bf 83}, 034013 (2011)
  [arXiv:1008.3632 [hep-ph]].

\bibitem{Guo:2009wr}
  F.~-K.~Guo, C.~Hanhart and U.~-G.~Mei{\ss}ner,
  Phys.\ Rev.\ Lett.\  {\bf 103}, 082003 (2009)  [Erratum-ibid.\  {\bf 104}, 109901 (2010)]  [arXiv:0907.0521 [hep-ph]].

\bibitem{Liu:2013vfa}
  X.~-H.~Liu and G.~Li,
  Phys.\ Rev.\ D {\bf 88}, 014013 (2013)
  [arXiv:1306.1384 [hep-ph]].


\bibitem{Guo:2013zbw}
  F.~-K.~Guo, C.~Hanhart, U.~-G.~Mei{\ss}ner, Q.~Wang and Q.~Zhao,
  Phys.\ Lett.\ B {\bf 725}, 127 (2013)
  [arXiv:1306.3096 [hep-ph]].


\bibitem{Wang:2013hga}
  Q.~Wang, C.~Hanhart and Q.~Zhao,
  Phys.\ Lett.\ B {\bf 725}, no. 1-3, 106 (2013)
  [arXiv:1305.1997 [hep-ph]].

\bibitem{Voloshin:2013ez}
  M.~B.~Voloshin,
  Phys.\ Rev.\ D {\bf 87}, no. 7, 074011 (2013)  [arXiv:1301.5068 [hep-ph]].

\bibitem{Voloshin:2011qa}
  M.~B.~Voloshin,
  Phys.\ Rev.\ D {\bf 84}, 031502 (2011)  [arXiv:1105.5829 [hep-ph]].

\bibitem{Guo:2010zk}
  F.~-K.~Guo, C.~Hanhart, G.~Li, U.~-G.~Meissner and Q.~Zhao,
  Phys.\ Rev.\ D {\bf 82}, 034025 (2010)
  [arXiv:1002.2712 [hep-ph]].

\bibitem{Chen:2011pu}
  D.~-Y.~Chen, X.~Liu and T.~Matsuki,
  Phys.\ Rev.\ D {\bf 84}, 074032 (2011)
  [arXiv:1108.4458 [hep-ph]].

\bibitem{Chen:2012yr}
  D.~-Y.~Chen, X.~Liu and T.~Matsuki,
  Chin.\ Phys.\ C {\bf 38}, 053102 (2014)
  [arXiv:1208.2411 [hep-ph]].

\bibitem{Chen:2013bha}
  D.~-Y.~Chen, X.~Liu and T.~Matsuki,
  Phys.\ Rev.\ D {\bf 88}, 014034 (2013)
  [arXiv:1306.2080 [hep-ph]].

\bibitem{Adachi:2012cx}
  I.~Adachi {\it et al.}  [Belle Collaboration],
  arXiv:1209.6450 [hep-ex].

\bibitem{Casalbuoni:1996pg}
  R.~Casalbuoni, A.~Deandrea, N.~Di Bartolomeo, R.~Gatto, F.~Feruglio and G.~Nardulli,
  Phys.\ Rept.\  {\bf 281}, 145 (1997)
  [arXiv:hep-ph/9605342].



\bibitem{Isola:2003fh}
  C.~Isola, M.~Ladisa, G.~Nardulli and P.~Santorelli,
  Phys.\ Rev.\  D {\bf 68}, 114001 (2003)
  [arXiv:hep-ph/0307367].

\bibitem{Colangelo:2003sa}
  P.~Colangelo, F.~De Fazio and T.~N.~Pham,
  Phys.\ Rev.\  D {\bf 69}, 054023 (2004)
  [arXiv:hep-ph/0310084].

\bibitem{Liu:2009vv}
  X.~H.~Liu and Q.~Zhao,
  Phys.\ Rev.\  D {\bf 81}, 014017 (2010)
  [arXiv:0912.1508 [hep-ph]].

\bibitem{Liu:2010um}
  X.~H.~Liu and Q.~Zhao,
  J.\ Phys.\ G {\bf 38}, 035007 (2011)
  [arXiv:1004.0496 [hep-ph]].

\bibitem{Li:1996yn}
  X.~Q.~Li, D.~V.~Bugg and B.~S.~Zou,
  Phys.\ Rev.\  D {\bf 55}, 1421 (1997).

\bibitem{Locher:1993cc}
  M.~P.~Locher, Y.~Lu and B.~S.~Zou,
  Z.\ Phys.\ A {\bf 347}, 281 (1994)  [nucl-th/9311021].

\bibitem{Li:1996cj}
  X.~-Q.~Li and B.~-S.~Zou,
  Phys.\ Lett.\ B {\bf 399}, 297 (1997)  [hep-ph/9611223].

\bibitem{Weinberg:1962hj}
  S.~Weinberg,
  Phys.\ Rev.\  {\bf 130}, 776 (1963).

\bibitem{Weinberg:1963zza}
  S.~Weinberg,
  Phys.\ Rev.\  {\bf 131}, 440 (1963).

\bibitem{Weinberg:1965zz}
  S.~Weinberg,
  Phys.\ Rev.\  {\bf 137}, B672 (1965).

\bibitem{Badalian:1981xj}
  A.~M.~Badalian, L.~P.~Kok, M.~I.~Polikarpov and Y.~.A.~Simonov,
  Phys.\ Rept.\  {\bf 82}, 31 (1982).

\bibitem{Guo:2008zg}
  F.~-K.~Guo, C.~Hanhart and U.~-G.~Meissner,
  Phys.\ Lett.\ B {\bf 665}, 26 (2008)
  [arXiv:0803.1392 [hep-ph]].

\end{thebibliography}
\end{document}